# Temporal Fidelity in Dynamic Social Networks

Arkadiusz Stopczynski[1,2], Piotr Sapiezynski[1], Alex 'Sandy' Pentland[2], Sune Lehmann[1,3]

[1]*Department of Applied Mathematics and Computer Science, Technical University of Denmark, Kgs. Lyngby, Denmark*

[2]*Media Lab, Massachusetts Institute of Technology, Cambridge, MA, USA*

[3]*The Niels Bohr Institute, University of Copenhagen, Copenhagen, Denmark*

**It has recently become possible to record detailed social interactions in large social systems with high resolution. As we study these datasets, human social interactions display patterns that emerge at multiple time scales, from minutes to months. On a fundamental level, understanding of the network dynamics can be used to inform the process of measuring social networks. The details of measurement are of particular importance when considering dynamic processes where minute-to-minute details are important, because collection of physical proximity interactions with high temporal resolution is difficult and expensive. Here, we consider the dynamic network of proximity-interactions between approximately 500 individuals participating in the Copenhagen Networks Study. We show that in order to accurately model spreading processes in the network, the dynamic processes that occur on the order of minutes are essential and must be included in the analysis.**

## 1 Introduction

Temporal networks provide an important framework for modeling a variety of real systems [1]. Examples of complex systems where dynamics can play a central role include social networks, energy grids, networks of sexual contacts, and transportation systems [2–8].

Only recently, thanks to technical developments in data collection, it has become possible to collect high-resolution data about physical and virtual interactions in complex social systems. Using sociometric badges or smartphones, it is now possible to record interactions happening on multiple channels and at multiple timescales, measuring events with minute-by-minute resolution [6,9–12]. With access to such data, we can begin to describe the complexity, structure, and dynamics of such social systems [13]. Accurate measurements and models of social systems are necessary in order to understand how diseases spread [6,7], what makes teams productive [14,15], or how friendships form and disappear [11,13].

A fully-formed framework for incorporating network dynamics has yet to be established [?,?,16–18]. We know, however, that for many practical applications, it is important to get the details right, because variations in how the time dimension is incorporated can lead to significant differences in the modeling of dynamical processes unfolding on the network. Understanding spreading in dynamic networks is of particular interest, as these may represent a wide variety of processes in the system, including spreading of biological pathogens, information, knowledge, or behaviors.

Recently, there has been a growing interest in how to correctly and efficiently incorporate time dimension in the modeling of disease spread. Over the last few years, studies have focused on the mixing matrices capturing important epidemiological features [?], efficient representation of the spreading networks with coarser temporal representation [?,16], and the fundamental impact that temporal features have on the spreading process [17,18]. Here we study how the fidelity of representation of network behavior at short timescales—on the order of minutes—influences simulated spreading in the network. These minute-to-minute dynamics are particularly interesting because data collection with high temporal resolution tends to be challenging and costly. We consider the ramifications of reducing temporal resolution and which biases such a reduction introduces in terms of understanding spreading process in the temporal network of close proximity interactions.

**The Dataset** Here we analyze close proximity interactions network of participants of the Copenhagen Networks Study (CNS) [9]. This proximity dataset is based on Bluetooth scans collected using state-of-the-art smartphones. We define an interaction between users $i, j$ in 5-minute timebin $t$ as $\gamma_{ijt} = s$, where the signal strength $s$ is reported by the handsets as received signal strength (RSSI). Since Bluetooth scans are unlikely to result in false positives, we use a symmetrized observations matrix (and resulting undirected network), assuming that $\gamma_{ijt}$ is present if $\gamma_{jit}$ exists. In this work we focus on how different sampling scenarios influence the overall results of spreading simulations. Our emphasis is not on investigating who would be infected in a real epidemic outbreak. In order to be able to measure the impact of subsampling, we need high resolution data as a reference. Therefore, we only use the data from participants with high data quality: out of 696 freshmen students active in February 2014, we select 476 participants with data quality of at least 60% (fraction of 5- minute bins in which the data is available). We understand that removing participants from the analysis might affect the network structure in a way that slows down the spreading processes. The resulting dataset is nevertheless the largest of its kind and, as we show in the following, represents a dense network.

## 2 Results

**Dynamics of a complex social system** The network of close proximity interactions in the CNS dataset displays dynamics at multiple time scales. We can observe distinct weekly and diurnal patterns (Figure 1a). Concurrently, at the minute-by-minute resolution, we observe significant fluctuations in the number of links active within 5-minute windows (Figure 1a inset).

We can quantify the magnitude of network changes by considering the overlap between active links in consecutive time slices as a function of the duration of the aggregation time window. A link $(i, j)$ is considered active if at least one interaction happened on it within the aggregated time window. We define the overlap as

$$J = \frac{|L_t^{\Delta T} \cap L_{t+1}^{\Delta T}|}{|L_t^{\Delta T} \cup L_{t+1}^{\Delta T}|} \quad (1)$$

where $L_t^{\Delta T}$ is set of links present in time $t$, in a time window of size $\Delta T$. The overlap, averaged over all time-bins in the network, is large at shortest timescales ($J(\Delta T = 5$ minutes$) = 0.71$) but drops rapidly as the size of the window increases (Figure 1b). For example, $J(\Delta T = 1\text{h}) = 0.51$, indicating a substantial turnover even at short timescales.

Within the 5-minute time-bins, the network is comprised of disjoint cliques, each one corresponding to a gathering of individuals. The changes of the network during short time intervals can be attributed to people moving between gatherings, as proposed by Sekara *et al.* [13]. As a system, these are constantly evolving, with members changing associations, and gatherings dissolving and forming (Figure 2a). These changes lead to the network connectivity that can be observed when aggregating the interactions into time-bins of longer duration (Figure 2b), even though every single time slice still consists of disjoint cliques.

**Temporal subsampling** To study the effect of temporal subsampling—reduction of dynamical information in the network—on the spreading processes, we consider two sampling schemes. These schemes are motivated by data collection strategies employed in the real studies, and therefore not necessarily an effort to devise the best possible temporal compression strategy.

In the first approach, which we call *snapshot sampling*, we choose a random 5-minute bin from every $N$ bins and consider this to represent the state of the entire network for these $N$ bins (Figure 3a). This way, we reduce the temporal resolution by a factor of $N$, but in a coherent fashion, because the network slices we use contain actual observed network states. The 'snapshot sampling' is typically the result of data collection methods which use static snapshots of the full population measured simultaneously. This is the case when physical proximity networks are inferred based on photographs or synchronized sensors, for example reports from a WiFi system (i.e. a list of devices connected to a router at the time of taking the snapshot, as in [15]), or results of Bluetooth scans performed by a fixed location device (as in [19,20]). We use random 5-minute bin from every $N$ bins rather than first bin (or last or middle) to remove the possible bias of choosing always the same part of the large timebin (for example, always choosing beginning of the hour).

In the second approach, which we refer to as *link sampling*, we sample the state (interacting/not interacting) of every link $(i, j)$ in the network from a random 5-minute bin within the sampling interval (Figure 3b). Thus, for every dyad, we choose a random 5-minute bin (out the $N$ eligible bins) use the dyad's state (on/off) as representative for this link in the subsampled network.

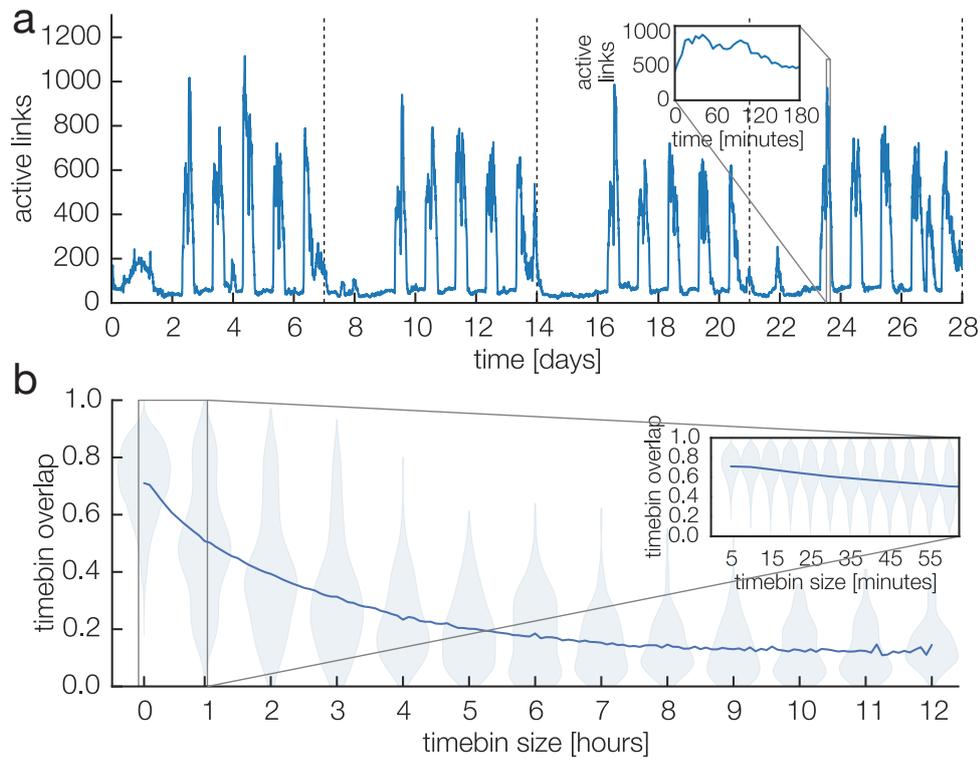

Figure 1: **Dynamics of network of close proximity in a complex social system.** (a) The network of proximity interactions displays distinct weekly and daily patterns. The number of active links can change drastically even within minutes, as show in the inset. (b) Mean overlap of active links between network slices as a function of aggregation time window. The overlap is high for short time windows but drops rapidly when longer windows are considered. The violin plot shows the exact distributions for chosen timebin sizes. The drop in the overlap most significant in the initial part of the aggregation (inset).

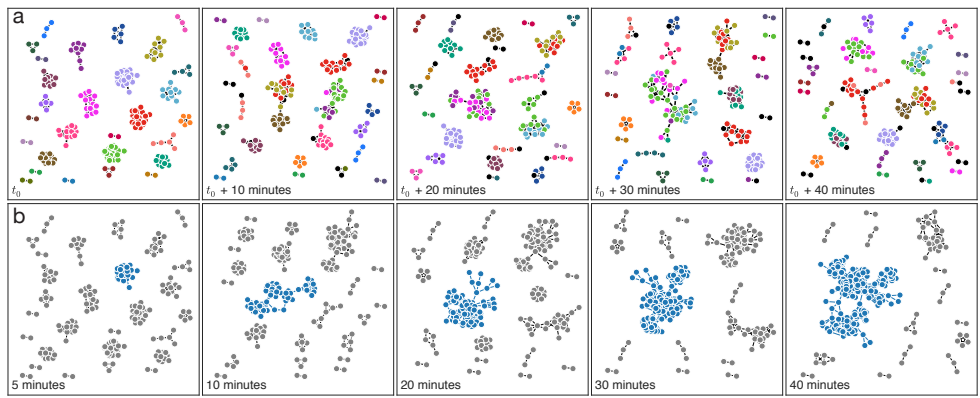

Figure 2: **Dynamics of physical proximity networks in a complex social system.** (a) Nodes are colored based on the component they belong at (randomly chosen) $t_0$ (5-minute timebin). While preserving the colors we plot the network in $t_2, t_4, t_6, t_8$, corresponding to $10, 20, 30, 40$ minutes later. Nodes that are not present at $t_0$ are marked in black. We can see how nodes move between gatherings. (b) The constant mixing of the nodes presented in (a) connects the initially separate components in to a well connected network when aggregating the interactions—even at relatively short timescales. Largest connected component in the network is highlighted.

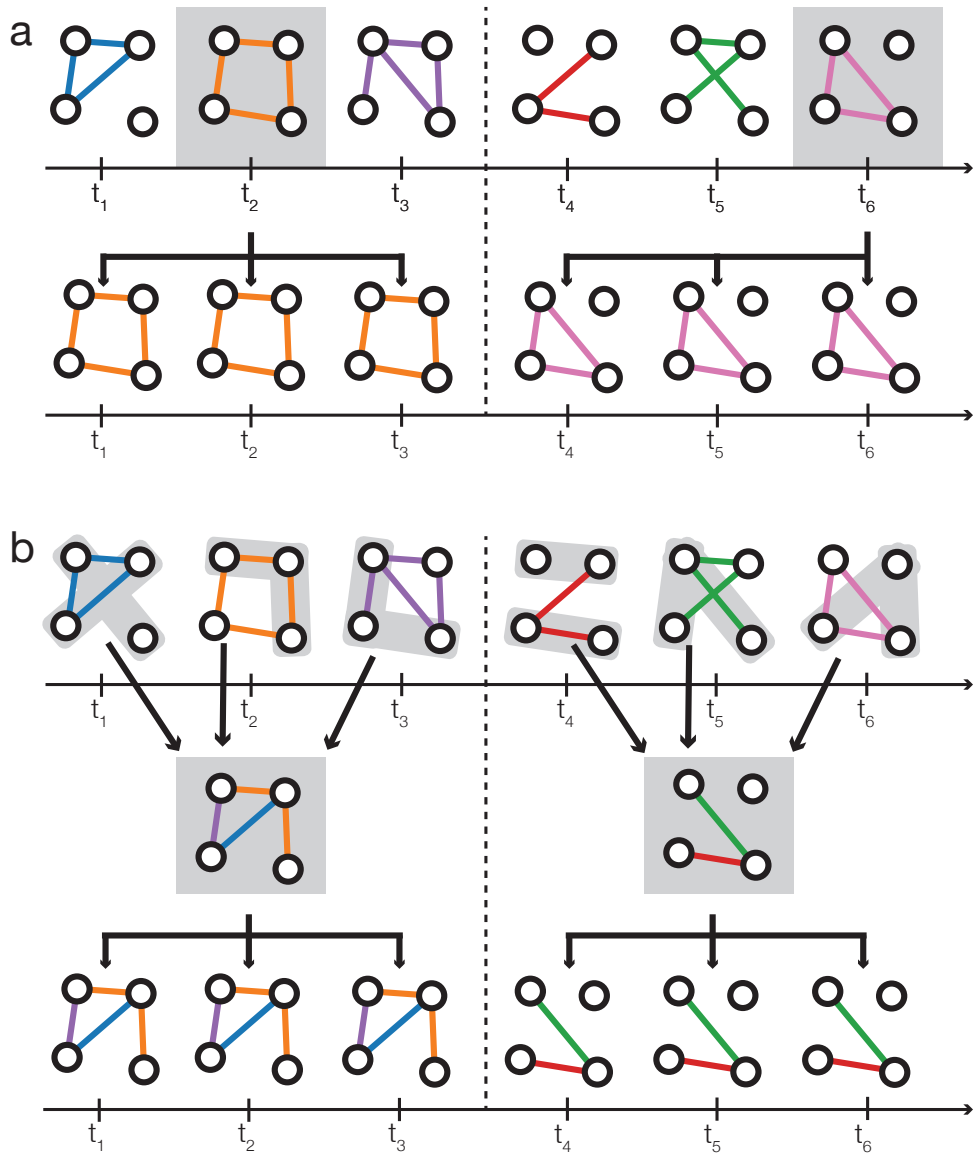

Figure 3: **Models for temporal subsampling.** (a) Snapshot sampling: when reducing the sampling resolution $N$ times, for each set of $N$ consecutive 5-minute bins we randomly select one and use it to represent these $N$ bins. (b) Link sampling: when reducing the sampling resolution $N$ times, for each dyad in the network during $N$ consecutive 5-minute bins we choose its state (interacting or not interacting) in a random bin among the $N$ and use the state to represent the dyad's status across the $N$ bins.

This sampling strategy results in a network state which may have never existed at any given point in time, but which also contains a factor of $N$ less temporal information compared to the original network. The link sampling corresponds to sampling occurring in multiple places in the population in an asynchronous way, a situation which occurs when collecting data using mobile phones or sociometric badges.

The temporal subsampling in both modes reduces the information about dynamics and results in a lossy compressed version of the temporal network. The information about the exact dynamics is lost (Figure 4a), replaced by static representations, with a width corresponding to the subsampling parameter $N$. In both subsampling scenarios the probability that a link $(i, j)$ is active is directly related to the number $n$ of 5-minute bins in the $N$-bin interval in which the link is present and this probability is equal to $n/N$. This implies that the average number of interactions after subsampling is the same in both snapshot and link sampling. However, due to the high temporal variability presented in the inset to Figure 1a, we expect that the number of links in the snapshot sampling will have a higher variation. This is, in fact, the case, as we show in Figure 4b. As expected, snapshot sampling results in a larger variability in the total number of interactions, because snapshots with very high or very low number of links may end up being chosen; still, the variability is within $\pm 20\%$ from the number of interactions for the full network. Performing linear regression on the mean values of total number of interactions we test for slope different from $0$ ($H_0 : b = 0$). The test statistic is $t = b/s_b$ on $(N - 2)$ degrees of freedom, and for both snapshot and link sampling we do not discover any significant trend in the average number of total interactions ($p = 0.60$ and $p = 0.63$ respectively). On average, when the number of interactions is considered, the subsampled networks are equal.

In spite of the fact that the total amount of temporal information and average number of total interactions are equal, the structure of the snapshot and link networks is quite different. Keeping track of the size of the largest connected component (LCC) we notice that the coherent network sampling results in disconnected neighborhoods dominating the network (Figure 4c). As expected, in link sampling, the network is more connected, with LCC containing up to $50\%$ of the nodes in the network.

**Spreading results** To quantify the effect of temporal subsampling on the modeling of a dynamic process unfolding on the network, we simulate spreading using a Susceptible-Infected-Recovered (SIR) model. In the spreading, we explore a variety of values for the transmission parameter $\beta$, including very slow and very fast transmissions (ranging between $\beta = 0.002$ and $\beta = 0.05$), and maintain a fixed recovery parameter $\mu = 4$ days. We randomly subsample the network 10 times for every value of subsampling parameter $N$ and run 100 simulations per condition, with a random starting time-bin and index patient. We apply circular boundary condition to extend the network beyond one month. For our purposes, the spreading simulation is used to understand the impact of sampling on a dynamic processes in the network. We do not attempt to model any particular disease.

Temporal subsampling, both snapshot and link-based, results in decreased spreading. The spreading process is slower, with a smaller peak value, and reduced total outbreak size (Figure 5a). This effect is more pronounced for rapid spreading (large $\beta$) and the effect is markedly stronger for snapshot sampling.

In Figure 5b, we quantify the effect of temporal subsampling on outbreak size. The drop of the outbreak size with the subsampling parameter $N$ is well explained by linear model (ordinary least squares regression), with a sub-linear effect for low values of $\beta$. Again, the effect is dramatically more pronounced for the snapshot subsampling. Similarly, probability of small outbreaks (reaching only a small fraction of the network) grows as a function of subsampling (Figure 5d), with effects much more pronounced for the snapshot sampling. Finally, a reduction of temporal fidelity drastically increases the time it takes for the spreading process to reach $50\%$ of the network (Figure 5e).

**Link-subsampling vs. full resolution** It is interesting to consider why spreading in the link-subsampled network is not faster than spreading in the full network: the number of connections in the link-subsampled network is typically higher than in single time slices of the full-resolution network. To understand why, we consider the structure of the link-subsampled network compared to the full-resolution network aggregated over the same time window. The difference arises from the fact that that although the subsampling is performed so that the subsampled and full network have the same number of interactions $(i, j, t)$ in any given time window (Figure 6a), the way these interactions are distributed on links $(i, j)$ is very different.

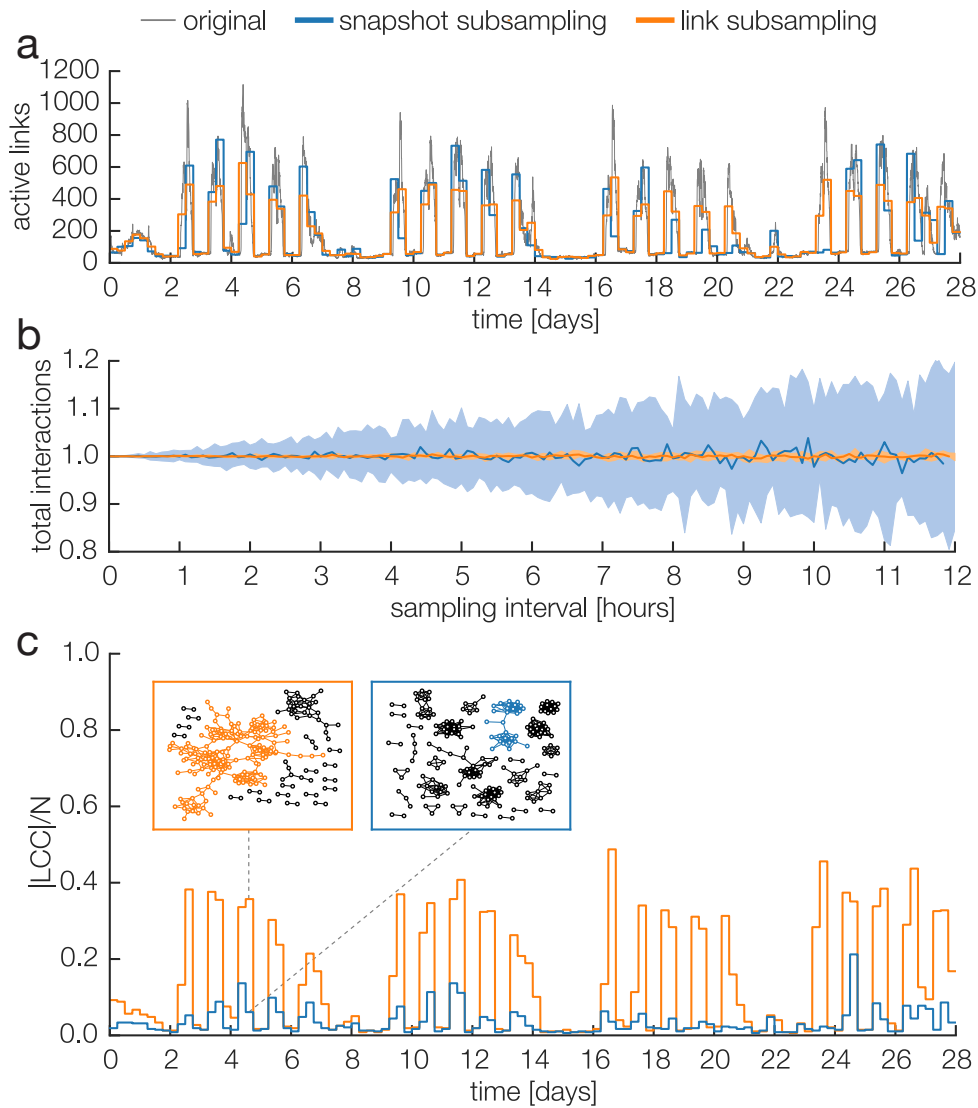

Figure 4: **Temporal subsampling of the network.** (a) The full dynamics are replaced by static representation, with the width depending on the subsampling parameter $N$. Here shown for $N = 72$, i.e. one sample per 6 hours. (b) Average number of total interactions in the network depending on the sampling parameter. Shaded areas indicate $10^{th}$ to $90^{th}$ percentiles across 100 simulations per time value. (c) Fraction of nodes contained in the largest connected component for one realization of the network subsampling ($N = 72$). Link sampling results in much more connected network.

To help guide our thinking about the differences, consider the full-resolution network aggregated over a certain time window. Here, the distribution of the link weights is broad, with many weak links and a few very strong connections. By contrast, link-subsampling creates a network where all links have the same weight—because all links are active through the entire window (Figure 6b). The full-resolution network has many more—but weaker—links active. This has strong implications for the connectedness of the network. In the link-sampled network, the network is split into a number of separate components, and an infection is rarely able to infect the entire network within a single frame. This is not the case in the full network, which has an effectively much larger connected component within each frame (Figure 6c). The way the full networks grows connected across time-slices is shown in Figure 2b.

**Slow versus fast spreading** These dynamics are sensitive to the speed of the disease spread. When the disease spreading is slower (low $\beta$) than the changes in the network, the gradual building of the connectivity in the full-resolution network does not slow down the spreading: from the slow disease perspective the network looks well connected. In the case when the disease spread is high (high $\beta$), the lower number of links $(i, j)$ in the link-subsampled network becomes the limiting factor. When the transmission parameter is large, the the number of links, not the link weights, is important. The disease is unable to reach the full network, for example gettting 'stuck' in a disconnected component. In both of these cases the full-resolution network facilities spreading, due to higher number of links.

These findings imply there may exist a third regime of $\beta$, where the transmission is faster than changes in the full network, 'waiting' for connectivity in the full-resolution network to build up, but slow enough that it does not run out of links in the subsampled network (never fully filling up its network components). Such regimes can be found in the network for fixed starting conditions (start time and node). But these cases are are rare, because each instance depends on an interplay between the structure of the network, size of the sampling window, and starting conditions. Thus, when averaged over many different starting conditions, the spreading is slower in the link-subsampled network due to the lack of the high number of weak links (Figure 6d). In the following, we discuss findings for the averaged case.

As expected, the impact of losing the temporal fidelity is strongest for fast spreading processes. With the lack of information about the detailed network dynamics, the disjoint gatherings produced by snapshot sampling lead to containment of the disease, resulting in smaller and slower spreading. When the transmission parameter $\beta$ is high (fast spreading), the disease is more likely to infect all nodes in the available neighborhoods, with no possibility to propagate to new places. For slow processes, the loss of temporal fidelity is less significant, as the spreading takes more time to fill up the isolated gatherings. The containment effect is much smaller in the link subsampling, as the network is more connected due to different (non-coherent) configuration of the links.

## 3 Discussion

Above we have investigated how modeling of spreading processes is impacted by reducing the temporal fidelity of close proximity interaction networks. We found that the network are highly dynamic, even at at short timescales. Within short time-bins, nodes gather in disjoint cliques, but with changing affiliation across time. These dynamics create significant interconnectedness when considering network at longer timescales (hours). When these short term dynamics are disregarded, either due to data collection process or data compression, spreading processes are strongly affected—as the temporal fidelity decreases, outbreaks become less frequent and smaller.

Interestingly, subsampling the network in a synchronized way (when the state of the entire network is sampled at once and repeated) has a much greater impact on the spreading results than when sampling is performed independently across links. This is because the disjoint gatherings that appear at shortest timescales inhibit the spreading process, when the minute-to-minute dynamics of nodes switching membership are lost. When we sample every link from an independently chosen time-slice the impact is much smaller, effectively approximating these short timescale dynamics.

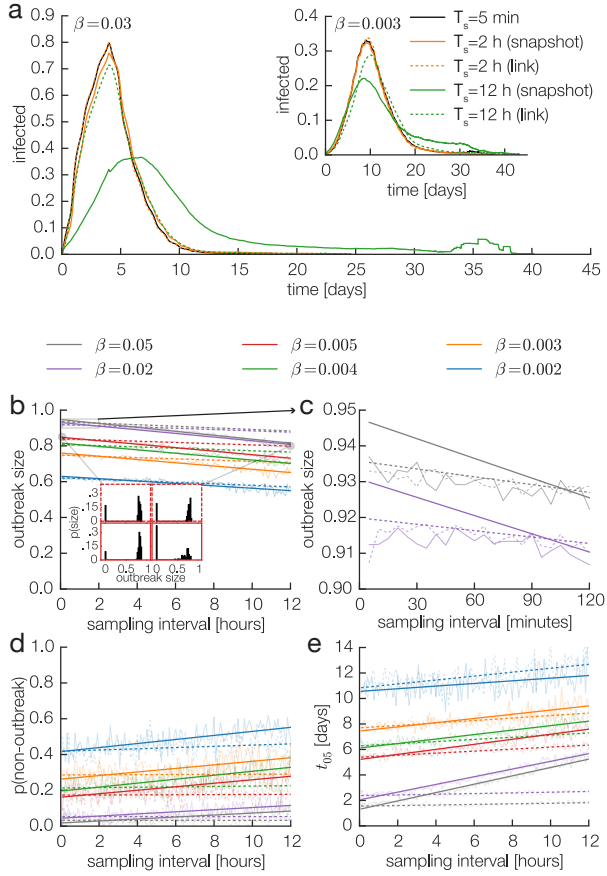

Figure 5: **Results of SIR spreading process.** (a) The shape of the spreading curve is less affected when each dyad is sampled independently, but network sampling leads to a significant underestimation of the outbreak size. The effect is less pronounced for slow spread (see the inset). Solid lines in (b)-(e) represent snapshot sampling, dashed lines represent link sampling. (b) With fast spreading epidemics, snapshot sampling results in a smaller expected size of the outbreak, but slow spreading epidemics are less affected. (c) Subsampling from $5$-minute bins down to $120$-minute bins (increasing the sampling interval $T_s$) does not significantly change the expected results of spreading simulations. (d) The probability of a non-outbreak (outbreak smaller than $20\%$ of the network) grows with the temporal subsampling. (e) For fast spreading epidemics, the time needed to infect half of the population grows linearly with the increasing subsampling rate in the network sampling scenario, but stays relatively stable when we sample dyads. Difference for low $\beta$ is not statistically significant between snapshot and link sampling.

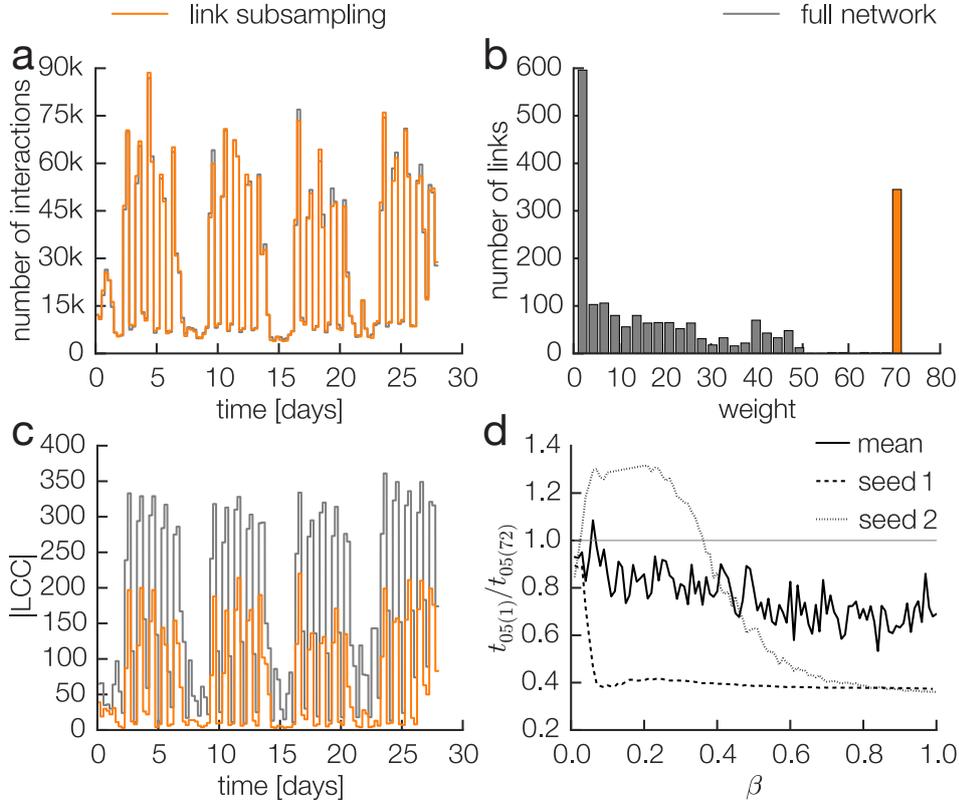

Figure 6: **Link weight heterogeneity in the full-resolution network.** (a) Links-subsampled and full-resolution network aggregated over the same time window have the same number of interactions $(i, j, t)$. Here shown for $N = 72$ i.e. $6h$ sampling. (b) The link weights in these views are distributed very differently, the full-resolution network features a long-tailed distribution with many weak links, whereas all links in the subsampled network have the same weight equal to sampling window size (72). (c) The higher number of links in the full-resolution network leads to a greater connectivity, here illustrated by the size of LCC. (d) The difference in the structure impacts the spreading. For fixed starting conditions (time-bin and seed node), it is possible to find regime of $\beta$ where the spreading on link-subsampled network is in fact faster (values above 1 on the plot). This is however not guaranteed for every starting condition and on average the spreading is slower in the link-subsampled network due to lower number of links.

The results presented here highlight a fundamental property of close proximity networks in social systems. We show how the dynamics contained within hourly time-bins can be instrumental for spreading process in the society. Simultaneously, from a methodological perspective, we illustrate how inclusion of these dynamics is crucial for understanding of the network of close proximity interactions and dynamical processes unfolding on them.

**Acknowledgements** We thank Vedran Sekara, and Yves-Alexandre de Montjoye for useful discussions. This research was funded, in part, by the Villum Foundation ("High Resolution Networks"), as well as the University of Copenhagen through the UCPH-2016 "Social Fabric" grant.

**Competing Interests** The authors declare that they have no competing financial interests.

**Correspondence** Correspondence and requests for materials should be addressed to Arkadiusz Stopczynski (email: arks@dtu.dk).